\documentclass[review]{elsarticle}

\usepackage{hyperref}

\makeatletter
\def\ps@pprintTitle{%
	\let\@oddhead\@empty
	\let\@evenhead\@empty
	\def\@oddfoot{\reset@font\hfil\thepage\hfil}
	\let\@evenfoot\@oddfoot
}
\makeatother

\begin{document}
	
\begin{frontmatter}

\title{Anomalous metallic oxygen band in the potential superconductor
	KCa$_2$Fe$_4$As$_4$O$_2$}

\author[addr_1,addr_2]{Nikita~S.~Pavlov}
\cortext[mycorrespondingauthor]{Corresponding author}
\ead[url]{pavlovns@lebedev.ru}

\author[addr_2]{Kirill.~S.~Pervakov}

\author[addr_1,addr_2]{Igor~A.~Nekrasov}

\address[addr_1]{Institute for Electrophysics, Russian Academy of
	Sciences, Ekaterinburg, 620016, Russia}
\address[addr_2]{P. N. Lebedev Physical Institute, Russian Academy of
	Sciences, Moscow, 119991, Russia}

\begin{abstract}
	The electronic structure, magnetism and Fermi surface of a hole
	self-doped potential iron-based superconductor KCa$_2$Fe$_4$As$_4$O$_2$
	(12442) were studied by a first-principes DFT/GGA approach.
	The projection onto Wannier functions basis is proposed to calculate directly the self-doping value. In particular for KCa$_2$Fe$_4$As$_4$O$_2$ it is 0.75 hole on Fe-3d states.
	Surprisingly, for the KCa$_2$Fe$_4$As$_4$O$_2$ it was found that except
	the Fe-3d bands the O-2p bands also cross the Fermi level, which is
	quite anomalous behavior for O-2p states.
	In general, the O-2p states are usually fully occupied in transition
	metal oxides.
	Here it is not the case due to the fact that CaO layer gives some of
	the electrons to the FeAs layer, thus leaving O-2p states partially occupied.
	The O-2p states form additional Fermi surface sheet with propeller-like
	shape around $\Gamma$-point and modify the Fe-3d Fermi surface sheets
	due to the hybridization. To prove the presence of anomalous metallic O-2p band in KCa$_2$Fe$_4$As$_4$O$_2$ the comparative study with the same family systems KCa$_2$Fe$_4$As$_4$F$_2$ and RbGd$_2$Fe$_4$As$_4$O$_2$ (where it is not the case) was performed.
	The magnetic ground state is found to be antiferromagnetic with checkerboard ordering.
	Therefore, we expect that the KCa$_2$Fe$_4$As$_4$O$_2$ might be a potential superconductor with unusual properties.
\end{abstract}


\end{frontmatter}

\section{Introduction}

The first discovery of superconductivity in 1144 (KCa$_2$Fe$_4$As$_4$~\cite{KCaFe4As4_cryst_tr_Iyo2016, KCaFe4As4_arpes_gap_Mou2016}) and 12442 fluorine systems (KCa$_2$Fe$_4$As$_4$F$_2$~\cite{KCa2Fe4As4F2_lattice_Wang2016}) was made in 2016.
After that, in 2017, the 12442 oxygen system was synthesized.
Experimentaly superconductivity was observed in RbGd$_2$Fe$_4$As$_4$O$_2$ ($T_c=35$~K)~\cite{RbGd2Fe4As4O2_exp}, in BaTh$_2$Fe$_4$As$_4$(N$_{0.7}$O$_{0.3}$)$_2$ ($T_c=30$~K)~\cite{BaTh2Fe4As4_NO2_exp} and in Rb$Ln_2$Fe$_4$As$_4$O$_2$ ($Ln = $Sm, Tb, Dy and Ho) ($T_c=34-36$~K)~\cite{RbLn2Fe4As4O2_lat_tr_mag_Wang2017}.
A whole range of compounds with oxygen $ALn_2$Fe$_4$As$_4$O$_2$ ($A = $K and Cs; $Ln = $lanthanides) was experimently investigated in the work~\cite{ALn2Fe4As4O2_exp} ($T_c=33-37$~K).

To our knowledge, there is just one single DFT study of 12442 oxygen system -- RbGd$_2$Fe$_4$As$_4$O$_2$~\cite{RbGd2Fe4As4O2_calc} and several DFT calculations of 12442 fluorine systems: KCa$_2$Fe$_4$As$_4$F$_2$ system~\cite{KCa2Fe4As4F2_calc_Wang2016,KCa2Fe4As4F2_calc_Singh2018}, KCa$_2$Fe$_4$As$_4$F$_2$ film~\cite{KCa2Fe4As4F2_calc_Li2020} and CsCa$_2$Fe$_4$As$_4$F$_2$~\cite{CsCa2Fe4As4F2_calc_Singh2018}.
Whereas several dozen of experimental papers are devoted to the 12442 fluorine compounds.
Let us list below those works that we are aware of.
The most studied system among them is KCa$_2$Fe$_4$As$_4$F$_2$ ($T_c=33.36$~K)~\cite{KCa2Fe4As4F2_mag_Smidman2018}.
Fermi surface, quasiparticle bands and superconducting gap are investigated by ARPES for the system in~\cite{KCa2Fe4As4F2_ARPES}.
A resistance and elastoresistance measurements were performed~\cite{KCa2Fe4As4F2_elast_resist_Terashima2020}.
The critical current densities $J_c$ through the KCa$_2$Fe$_4$As$_4$F$_2$ wire at 4.2~K is 10~kA/cm$^2$ under self-field, and 1~kA/cm$^2$ at 100~kOe~\cite{KCa2Fe4As4F2_wire}.
Magnetic susceptibility and penetration depth are reported in~\cite{KCa2Fe4As4F2_mag_Smidman2018}.
The anisotropy parameter $\gamma$ and London penetration depth $\lambda$ were obtained from vortex torque~\cite{KCa2Fe4As4F2_torque_Yu2019}.
An extremely high upper critical field ($H_{c2}(T) \sim 9$~T) for the fluorine compound was observed in~\cite{KCa2Fe4As4F2_Hc2_Wang2019}.
Strong Pauli paramagnetic effect in the upper critical field was reported in~\cite{KCa2Fe4As4F2_Hc2_Wang2020}.
An inelastic neutron scattering study on the low-energy spin excitations was done in the Ref.~\cite{KCa2Fe4As4F2_neutron_Hong2020}.
A systematic study of electrical resistivity, Hall coefficient, magneto-optical imaging, magnetization, and scanning transmission electron microscopy (STEM) analyses of single crystals was performed in the work~\cite{KCa2Fe4As4F2_systematic_Pyon2020}.
Temperature-pressure phase diagram was investigated in~\cite{KCa2Fe4As4F2_press_Wang2019}.
Low-temperature specific heat (SH) is measured~\cite{KCa2Fe4As4F2_tr_Wang2020}.
Nuclear magnetic resonance (NMR) study was done~\cite{KCa2Fe4As4F2_NMR_Wang2017,KCa2Fe4As4F2_NMR_Luo2020}.

The correlations between $T_c$ and structural parameters was measured and discussed for $A$Ca$_2$Fe$_4$As$_4$F$_2$ ($A = $Rb, Cs) compounds~\cite{ACa2Fe4As4F2_lat_tr_mag_Wang2017}.
A transport and magnetic properties of the cobalt substitution in the intrinsically hole-doped KCa$_2$(Fe$_{1-x}$-Co$_x$)$_4$As$_4$F$_2$ are investigated in~\cite{KCa2FeCo4As4F2_tr_mag_Ishida2017}.
A transport and magnetic measurements in single crystals of CsCa$_2$Fe$_4$As$_4$F$_2$~were performed in~\cite{CsCa2Fe4As4F2_tr_mag_Wang2019}.
The thermal conductivity, resistivity, upper critical field $H_{c2}(T)$ of CsCa$_2$Fe$_4$As$_4$F$_2$ were investigated in~\cite{CsCa2Fe4As4F2_tr_mag_Huang2019}.
The vortex phase diagram of RbCa2Fe4As4F2 via magneto-transport and magnetization measurements was performed in~\cite{RbCa2Fe4As4F2_Xing_2020}.

In this paper we investigate the electronic structure, magnetism and Fermi surface of a hole self-doped potential iron-based superconductor KCa$_2$Fe$_4$As$_4$O$_2$ within DFT/GGA approach.
The Wannier functions analysis is proposed to obtain the self-doping value and the value was found to be 0.75 hole per Fe ion.
Surprisingly, the O-2p band crosses the Fermi level and has a metallic behavior.
The O-2p states occupancy per O ion is about 5.5 (instead of fully occupied value of 6) which is also obtained form Wannier functions analysis.
To demonstrate the presence of anomalous metallic O-2p band in KCa$_2$Fe$_4$As$_4$O$_2$ the comparative study of its band structure with KCa$_2$Fe$_4$As$_4$F$_2$ and RbGd$_2$Fe$_4$As$_4$O$_2$ systems was performed.

\section{Computational details}
Since the KCa$_2$Fe$_4$As$_4$O$_2$ compound has not been synthesized yet, we use the space group and lattice parameters for KCa$_2$Fe$_4$As$_4$F$_2$~\cite{KCa2Fe4As4F2_lattice_Wang2016}.
The crystal structure of KCa$_2$Fe$_4$As$_4$F$_2$ has the tetragonal symmetry with I4/mmm space group at room temperature and the lattice constants $a=b=3.8684$~\AA\ and $c=31.007$~\AA~\cite{KCa2Fe4As4F2_lattice_Wang2016}.
\begin{table}
\caption{The optimized ion positions of paramagnetic KCa$_2$Fe$_4$As$_4$O$_2$. The ion positions of KCa$_2$Fe$_4$As$_4$F$_2$ from~\cite{KCa2Fe4As4F2_lattice_Wang2016} are shown in brackets. dFe-As1 and As1-Fe-As1 angle}
\begin{tabular}{lcccc} 
	atom & site & $x$ & $y$ & $z$ \\
	K & 2a & 0 & 0 & 0\\
	Ca & 4e & 0.5 & 0.5 & 0.21583 (0.2085) \\
	Fe & 8g & 0 & 0.5 & 0.10784 (0.1108) \\
	As1 & 4e & 0.5 & 0.5 & 0.06604 (0.0655) \\
	As2 & 4e & 0 & 0 & 0.14880 (0.1571) \\
	O (F) & 4d & 0 & 0.5 & 0.25
\end{tabular}
\label{crystal_positions}
\end{table}
The DFT calculations were performed within the full-potential linearized augmented plane-wave (FP-LAPW) code WIEN2k~\cite{WIEN2k2020} with Perdew-Burke-Ernzerhof generalized gradient approximation of exchange correlation functional (GGA)~\cite{PBE_1996}.
The ion positions were DFT optimized for KCa$_2$Fe$_4$As$_4$O$_2$ starting with those for KCa$_2$Fe$_4$As$_4$F$_2$ (see Table~\ref{crystal_positions}).
The Brillouin zone k-points grid was taken to be $16 \times 16 \times 16$.

For DFT calculation of KCa$_2$Fe$_4$As$_4$F$_2$ the experimental ion positions were used~\cite{KCa2Fe4As4F2_lattice_Wang2016} which are shown in brackets in Table~\ref{crystal_positions}.
The Quantum Espresso~\cite{QE_Giannozzi2009} was used for calculation of RbGd$_2$Fe$_4$As$_4$O$_2$ with the Gd-4f states considered as a part of the core states.
The projected augmented wave pseudopotentials (PAW) were applied.
The ion positions of RbGd$_2$Fe$_4$As$_4$O$_2$ were taken from~\cite{RbGd2Fe4As4O2_calc}.

To calculate shell and orbital occupancies the DFT/GGA Hamiltonian was projected onto Fe-3d, As-4p (O-2p) Wannier functions. The obtained Wannier-projected dispersions perfectly coincide with the DFT ones.
Wannier functions are obtained within Wannier90 code~\cite{Wannier90_Pizzi2020}.

\section{Results and discussion}
\begin{figure}[h]
	\includegraphics[width=0.9\linewidth]{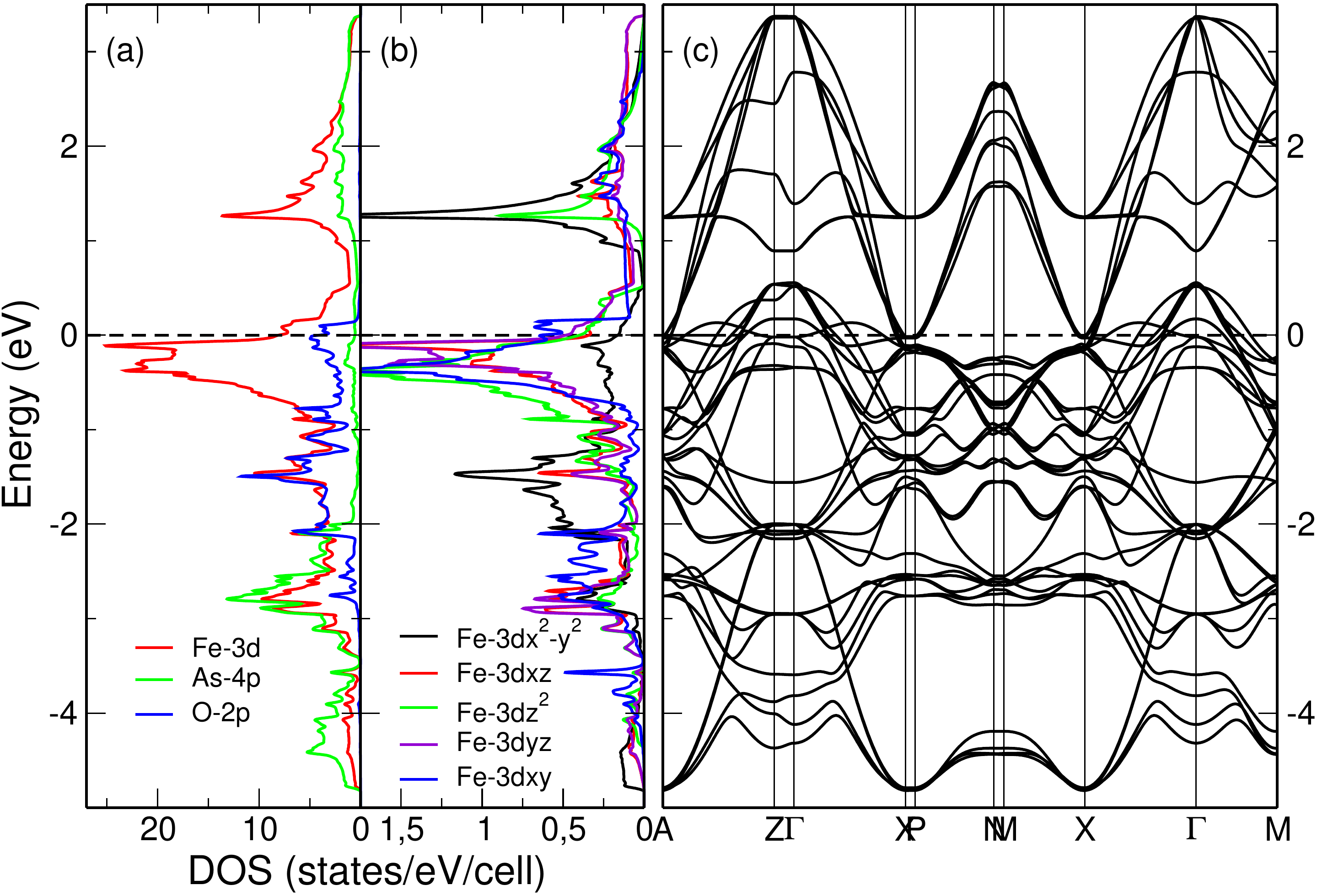}
	\caption{DFT/GGA (a) densities of states (Fe-3d, As-4p, O-2p), (b) densities of Fe-3d states and (c) band dispersions of paramagnetic KCa$_2$Fe$_4$As$_4$O$_2$.
		The Fermi level is at zero energy.}
	\label{LDA_dos_bands}
\end{figure}

The GGA densities of states (total and partial) and band dispersions of paramagnetic KCa$_2$Fe$_4$As$_4$O$_2$ are shown in Fig.~\ref{LDA_dos_bands}.
As usual for the iron-based superconductors the main contribution to the density of states at the Fermi level comes from Fe-3d states.
But in this compound surprisingly the O-2p states (which are located from $-2.8$~eV to $0.12$~eV) appear at the Fermi level as well.
It is quite unusual, anomalous behavior, because typically the O-2p states are fully occupied and lie in the energy interval from $-6$~eV to $-2$~eV.
It means that in KCa$_2$Fe$_4$As$_4$O$_2$ system the O-2p states of CaO layer do not receive enough electrons to get filled completely.
Thus the CaO layer gives some of the electrons to the FeAs layer.
The O-2p states at the Fermi level are mostly hybridized with the Fe-3d$_{xy}$ orbital.

There is the sharp peak of Fe-3d$_{xz}$ and Fe-3d$_{yz}$ states just below the Fermi level, which may be important for superconductivity in case of the hole doping.
The Fe-3d$_{xz}$ and Fe-3d$_{yz}$ bands become non-degenerate because of the different $z_{\rm As}$ distances below and above Fe ion plane since As has different neighbors there Ca or K.
The contribution to the density of states of Fe-3d($xz,yz,z^2$) orbitals is nearly the same at the Fermi level, while the contribution of Fe-3d$_{xy}$ is two times lager than later ones and Fe-3d$_{x^2-y^2}$ -- three times smaller.
The As-4p states are located in interval from $-4.8$~eV to $-2$~eV.

\begin{figure}[h]
	\includegraphics[width=0.6\linewidth]{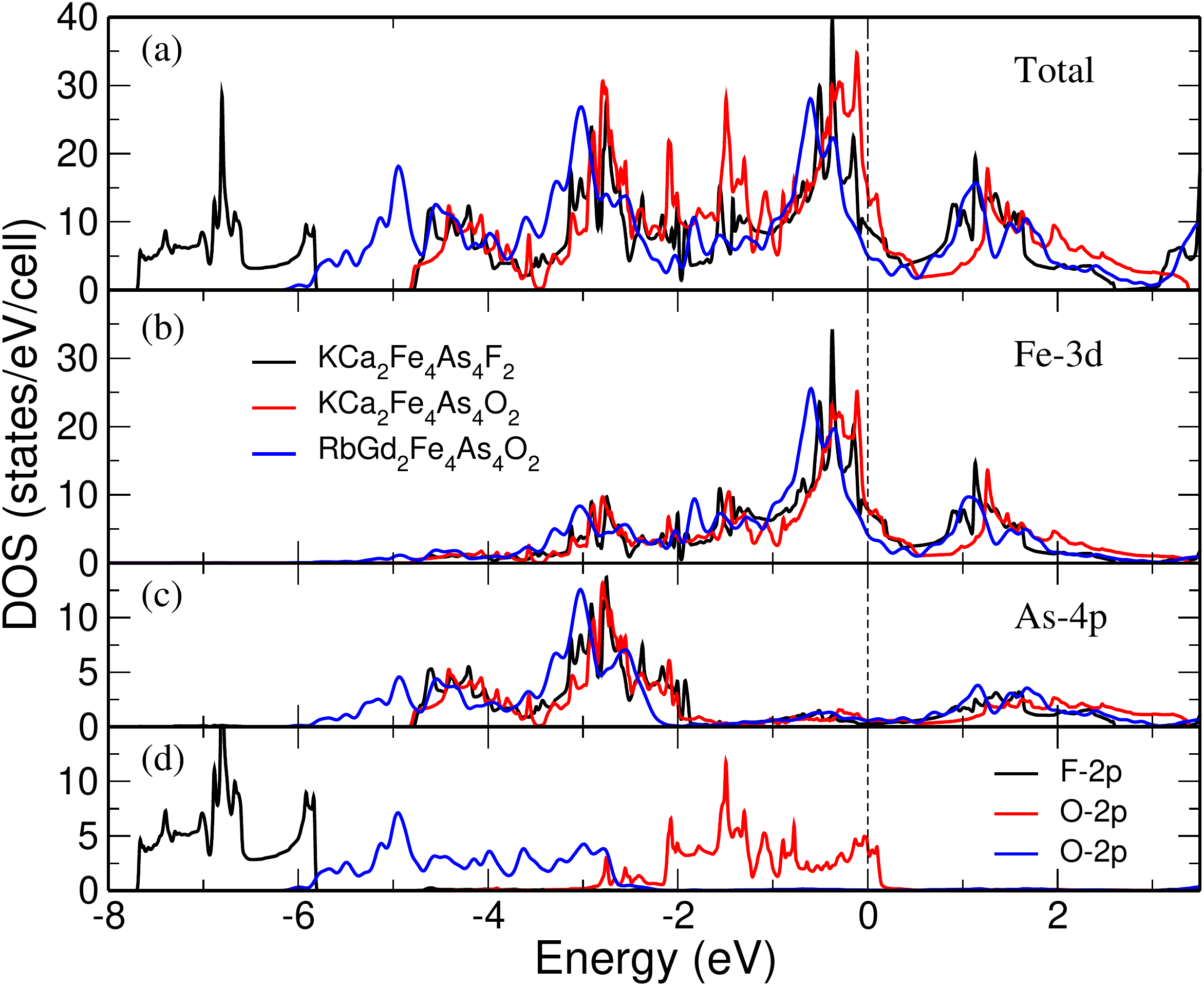}
	\caption{Comparison of DFT/GGA densities of states of KCa$_2$Fe$_4$As$_4$O$_2$ (red lines), KCa$_2$Fe$_4$As$_4$F$_2$ (black lines) and RbGd$_2$Fe$_4$As$_4$O$_2$ (blue lines): (a) total, (b) Fe-3d states, (c) As-4p states, (d) O-2p (F-2p) states.}
	\label{comapre_F2_RbGd}
\end{figure}

We compare the densities of states for KCa$_2$Fe$_4$As$_4$O$_2$ with ones of the RbGd$_2$Fe$_4$As$_4$O$_2$ to show the difference of the O-2p states energy position between them (see Fig.~\ref{comapre_F2_RbGd}).
For complitness
the densities of states of fluorine KCa$_2$Fe$_4$As$_4$F$_2$ system is also presented on Fig.~\ref{comapre_F2_RbGd}.
In KCa$_2$Fe$_4$As$_4$F$_2$ and RbGd$_2$Fe$_4$As$_4$O$_2$ systems the F-2p and O-2p states are fully occupied and are located well below the Fermi level.
The Fe-3d and As-4p states are similar to each other in shape and energy position for all three systems.
It should be noted that among all considered systems the KCa$_2$Fe$_4$As$_4$O$_2$ system has the largest value of density of states at the Fermi level.
Although the Fe-3d density of states at the Fermi level is equal to those of KCa$_2$Fe$_4$As$_4$O$_2$ and KCa$_2$Fe$_4$As$_4$F$_2$, the difference comes due to the O-2p states presence at the Fermi level.

\begin{figure}[h]
	\includegraphics[width=0.6\linewidth]{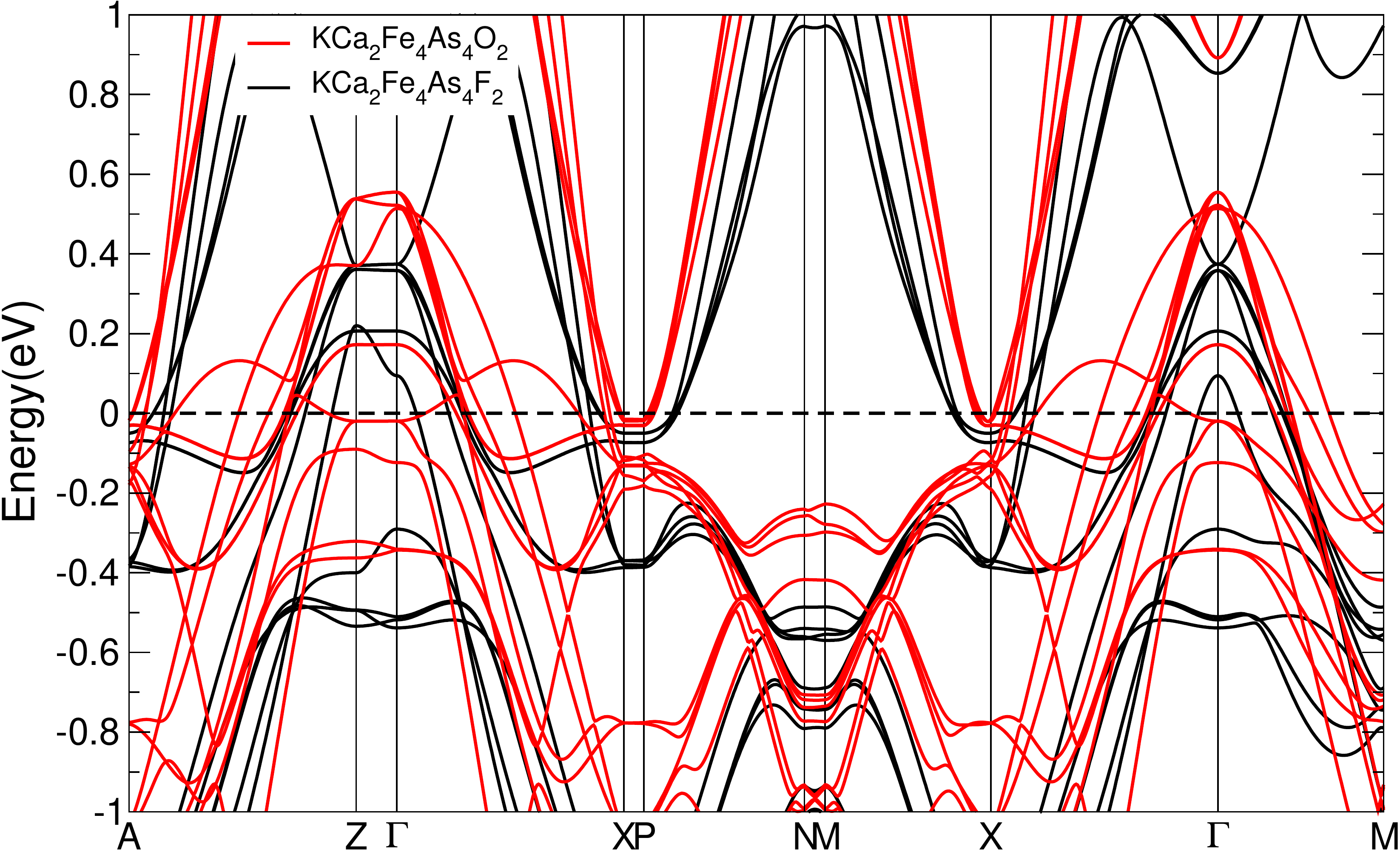}
\caption{Comparison of DFT/GGA band dispersions of paramagnetic KCa$_2$Fe$_4$As$_4$O$_2$ (red lines) with KCa$_2$Fe$_4$As$_4$F$_2$ system (black lines), which has two more electrons per formula unit.}
\label{comapre_F2_bands}
\end{figure}

\begin{figure}[h]
	\includegraphics[width=1.0\linewidth]{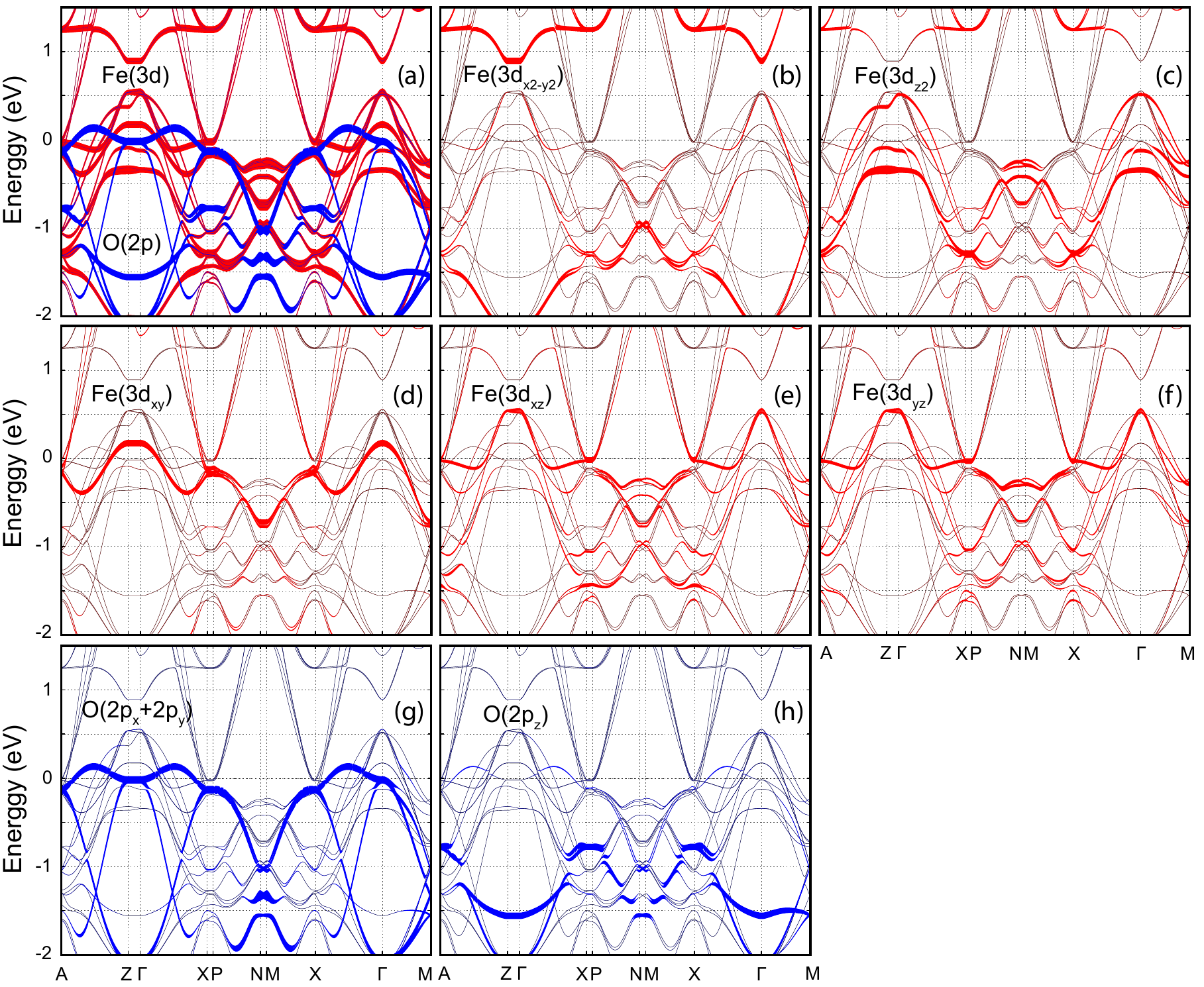}
	\caption{The projected band structure of paramagnetic KCa$_2$Fe$_4$As$_4$O$_2$, where the linewidth corresponds to the projected weight of Bloch states onto the (a) Fe-3d (red lines) and O-2p (blue lines), (b) Fe-3d$_{x^2-y^2}$, (c) Fe-3d$_{z^2}$, (d) Fe-3d$_{xy}$, (e) Fe-3d$_{xz}$, (f) Fe-3d$_{yz}$, (g) O-2p$_x$+2p$_y$, (h) O-2p$_z$ states.}
	\label{bands_orbs}
\end{figure}

Let's compare the band dispersions of KCa$_2$Fe$_4$As$_4$O$_2$ (red lines) and KCa$_2$Fe$_4$As$_4$F$_2$ (black lines) near the Fermi level on Fig.~\ref{comapre_F2_bands}.
In general, KCa$_2$Fe$_4$As$_4$F$_2$ bands (black lines) are about 0.1-0.2~eV lower in energy than KCa$_2$Fe$_4$As$_4$O$_2$ bands (red lines) since the fluorine system has two more electrons than the oxygen one.
Whereas, some bands have similar positions.
However, in KCa$_2$Fe$_4$As$_4$O$_2$ system there is additional band in the middle of $\Gamma$-X (A-Z) direction, which crosses the Fermi level.
This band is formed by the O-2p$_x$ and O-2p$_y$ orbitals as can be seen on Fig.~\ref{bands_orbs}(g).

Quite important question is how to calculate the self-doping value in the KCa$_2$Fe$_4$As$_4$O$_2$ system.
In general, in the parent iron-based compounds (for example, NaFeAs, BaFe$_2$As$_2$, LaOFeAs) valance of Fe is 2+ which corresponds to 6 electrons on Fe-3d shell and empty Fe-4s shell.
The self-doping means that for the stoichiometric system Fe-3d has less or more than 6 electrons and the valence of Fe is different from 2+.
One can calculate the formal valence of Fe in KCa$_2$Fe$_4$As$_4$O$_2$ by assuming the valencies of K$^+$, Ca$^{2+}$, As$^{3-}$ and O$^{2-}$ are known. As the result the valence of Fe will be 2.75+ (0.75 hole/Fe-3d).

The simplest way to estimate the self-doping value from DFT is just to take the values of DFT occupancies for instance in this work they are 6.12 for Fe-3d and 2.92 for As-4p.
But the As-4p states are located below the Fermi level (Fig.~\ref{LDA_dos_bands}), therefore, they are fully occupied and must have 6 electrons.
The reason of the discrepancy in DFT As-4p occupancy is that the DOS of As-4p states is spread in energy due to the hybridization with other states including strong hybridization with Fe-3d states.
In DFT, the total number of electrons is strictly fixed, but it can be difficult to calculate the partial contributions due to the hybridization.
To overcome this circumstance, we propose to use Wannier function projection, since  Wannier functions are maximally localized and have atomic-like nature.
It allows one to exclude the ``parasitic'' hybridization.

If it could be possible to disentangle the Fe-3d and As-4p bands and project only onto Fe-3d states,  then one could immediately obtain a filling of the Fe-3d shell.
In case of KCa$_2$Fe$_4$As$_4$O$_2$, this cannot be done with satisfactory agreement between DFT and Wannier function projected bands.
Therefore, we chose the Wannier function basis consisting of Fe-3d and As-4p orbitals (4 Fe atoms and 4 As atoms) and obtain the total occupancy 44.96 ($\approx 45$).
In general, the Fe-3d and As-4p states together have 12 electrons (6 electrons on Fe-3d and 6 electrons on As-4p), consequently the Fe$_4$As$_4$ block must have 48 electrons.
It means in KCa$_2$Fe$_4$As$_4$O$_2$ there are 3 electrons less per Fe$_4$As$_4$ block.
Assuming that the As-4p is filled completely the Fe-3d states must 5.25 electrons instead of 6.
This occupancy is very close to half-filling.
Thus, one can expect that Fe-3d electron-electron correlations might be noticeable.
From Fe-3d filling one can calculate the valance of Fe, assuming that the Fe-4s states are fully empty.
Hence the total number of holes on Fe can be found from sum $0.75+2.0=2.75$.

As mentioned above in KCa$_2$Fe$_4$As$_4$O$_2$ the O-2p states cross the Fermi level and thus are not fully occupied.
We include the O-2p orbitals into Wannier function basis set and obtain the total occupancy of 55.96 ($\approx 56$) for 4 Fe atoms, 4 As atoms and 2 O atoms.
Without self-doping this occupancy should be 60 (4* 6 electrons on Fe-3d, 4* 6 electrons on As-4p and 2* 6 electrons on O-2p).
Considering the result of projection to Fe-3d and As-4p orbitals, which have 3 electrons less, the O-2p states have 1 hole per 2 O atoms or 0.5 hole/O.

Let us perform the same analysis for KCa$_2$Fe$_4$As$_4$F$_2$ compound.
In case of Wannier function basis is set to Fe-3d, As-4p (F-2p) orbitals the total occupancy after the projection is 46.39 (58.36).
The F-2p states add another 12 electrons.
Therefore, the F-2p state are fully occupied and have 6 electrons.
While, the Fe$_4$As$_4$ block has 1.6 electrons less (0.4 hole/Fe-3d).
The formal valence of Fe in KCa$_2$Fe$_4$As$_4$F$_2$ with F$^{-}$ valence will be 2.25+ (0.25 hole/Fe-3d).
For the KCa$_2$Fe$_4$As$_4$F$_2$ the Wannier function projection and formal valence calculation provide different values.
It turns out that Fe get less electrons than expected from formal electroneutrality.

The projected bands of paramagnetic KCa$_2$Fe$_4$As$_4$O$_2$ onto the (a) Fe-3d (red lines) and O-2p (blue lines), (b) Fe-3d$_{x^2-y^2}$, (c) Fe-3d$_{z^2}$, (d) Fe-3d$_{xy}$, (e) Fe-3d$_{xz}$, (f) Fe-3d$_{yz}$, (g) O-2p$_x$+2p$_y$, (h) O-2p$_z$ states are presented in Fig.~\ref{bands_orbs}.
As can be seen from the density of states (Fig.~\ref{LDA_dos_bands}), the contribution of Fe-3d$_{x^2-y^2}$ orbital is quite far away from the Fermi level.
All other Fe-3d orbitals are located in the vicinity of the Fermi level.
The O-2p$_x$+2p$_y$ bands cross the Fermi level and are located between $-2.8$ and $0.12$ eV.

\begin{figure}[h]
	\includegraphics[width=1.0\linewidth]{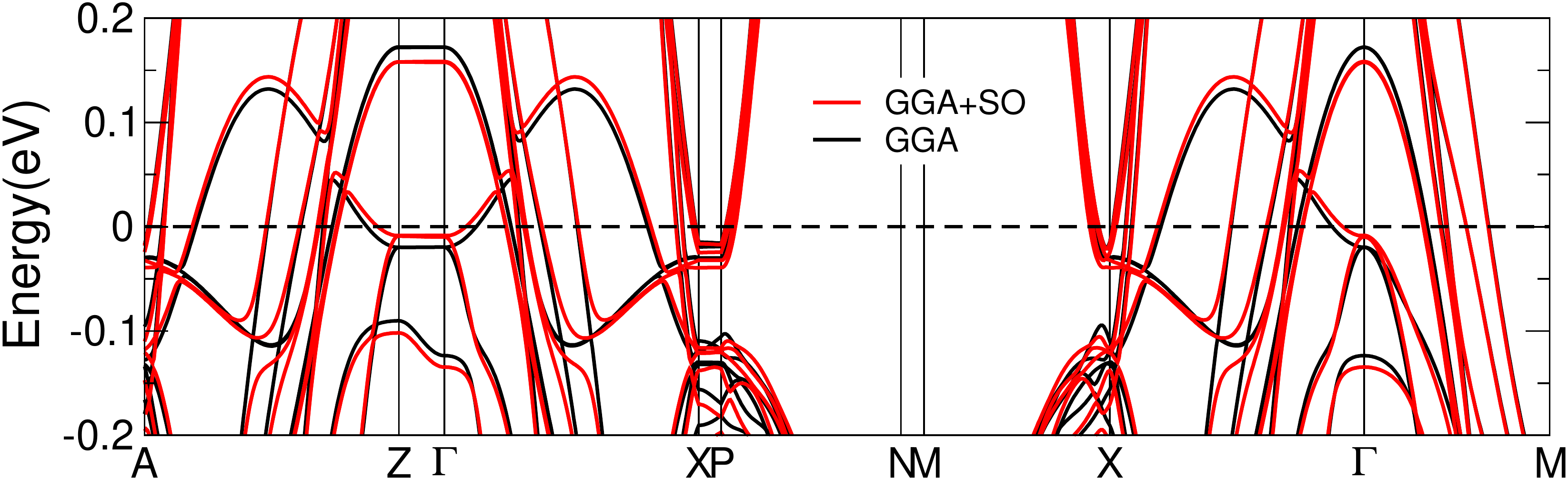}
	\caption{DFT/GGA band dispersions of paramagnetic KCa$_2$Fe$_4$As$_4$O$_2$ with spin-orbit coupling (SO) (red lines) and without SO (black lines).}
	\label{bands_SO}
\end{figure}

In the LDA+DMFT investigation of 1144 system KCaFe$_4$As$_4$~\cite{CaKFe4As4_Liu2020} it was shown that the spin-orbit (SO) coupling is important due to glide-mirror symmetry breaking.
We compare the band dispersion with and without SO near the Fermi level in Fig.~\ref{bands_SO}.
The SO gives rather small bands shift $\sim 0.01$~eV and bands spliting in A-Z, $\Gamma$-X directions around $-0.11$~eV.
Thus one can note, that the SO almost does not affect on the band structure of KCa$_2$Fe$_4$As$_4$O$_2$.

\begin{figure}[h]
	\includegraphics[width=1.0\linewidth]{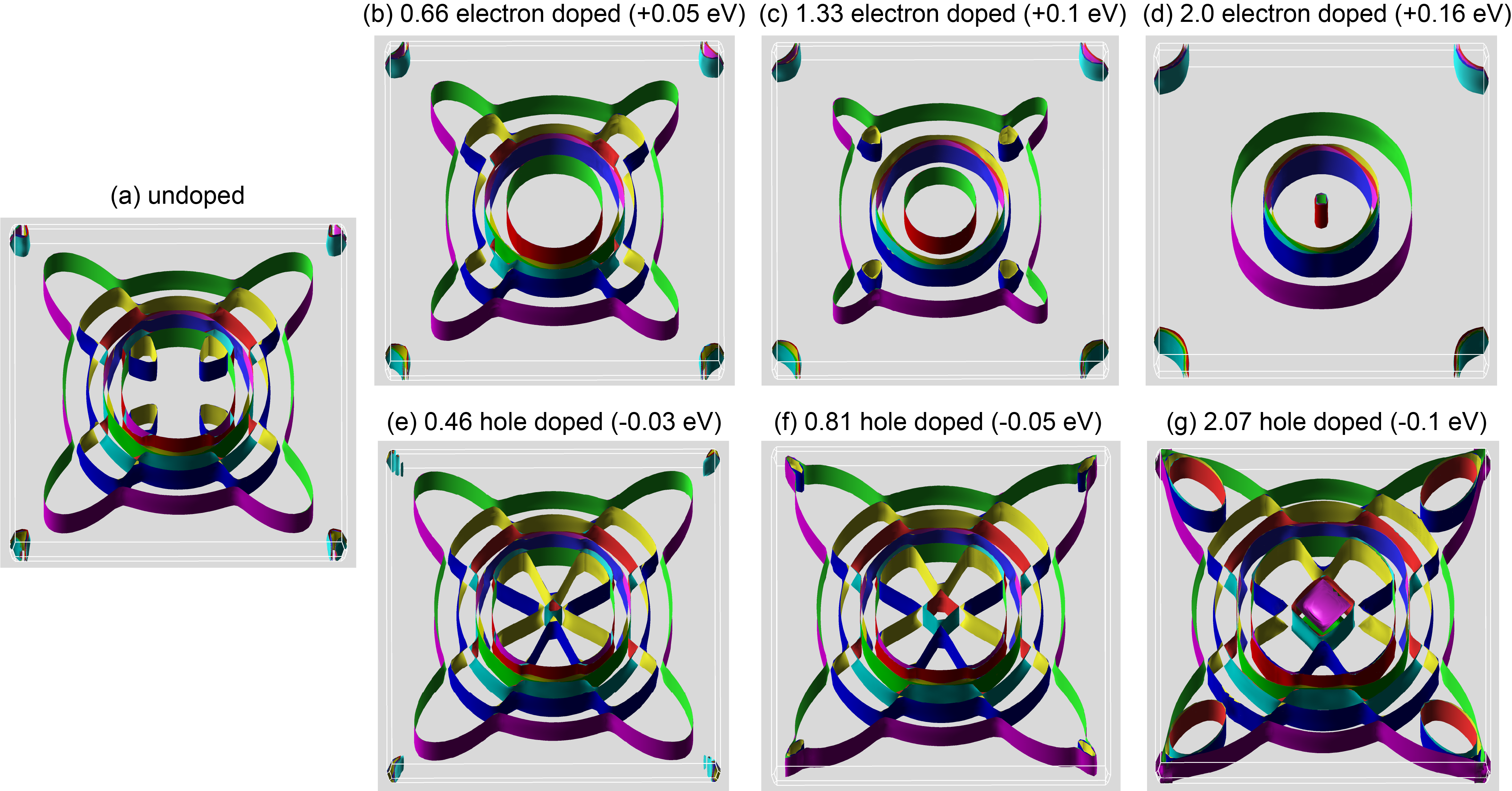}
	\caption{The Fermi surface of paramagnetic KCa$_2$Fe$_4$As$_4$O$_2$: (a) undoped case, (b) 0.66 electron doped (+0.05 eV rigid shift of Fermi level), (c) 1.33 electron doped (+0.1 eV shift), (d) 2.0 electron doped (+0.166 eV shift), (e) 0.46 hole doped (-0.03 eV shift), (f) 0.81 hole doped (-0.05 eV shift), (g) 2.07 hole doped (-0.1 eV shift).}
	\label{FS}
\end{figure}

In Fig.~\ref{FS} the DFT calculated Fermi surface for KCa$_2$Fe$_4$As$_4$O$_2$ system is preseted.
The Fermi surface sheet (Fig.~\ref{FS}a) with propeller shape around $\Gamma$-point corresponds to O-2p states.
Due to hybridization the typically cylindrical in iron-based superconductors Fe-3d Fermi surface sheets here are modified.

\begin{figure}[h]
	\includegraphics[width=1.0\linewidth]{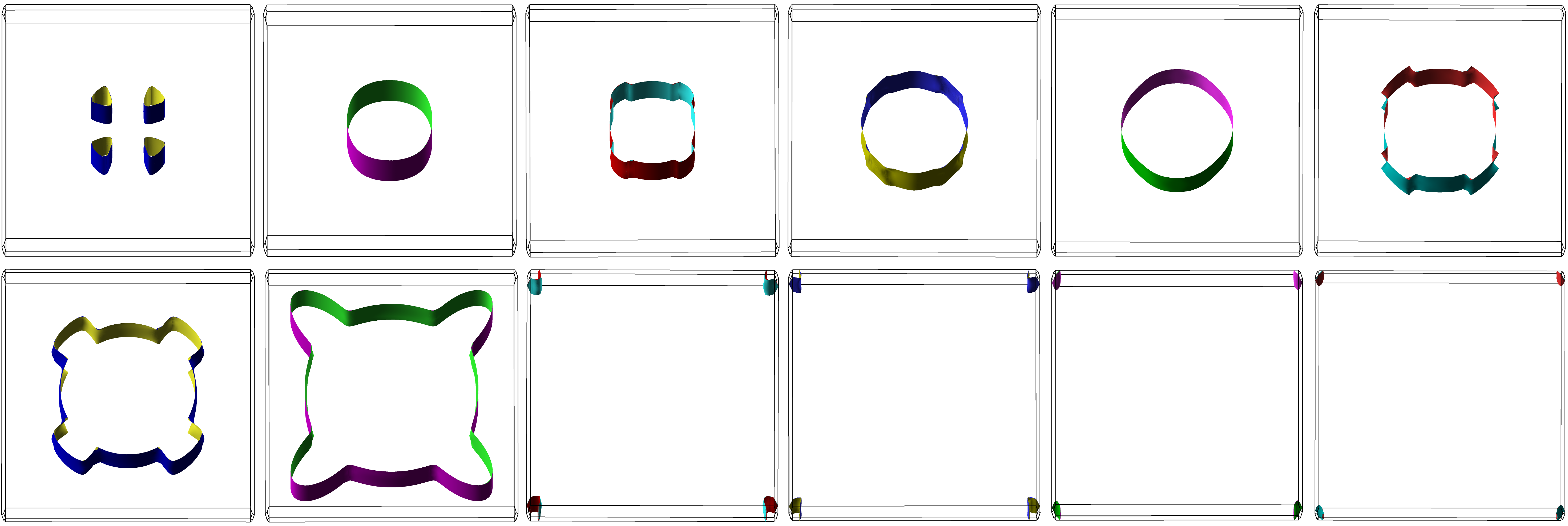}
	\caption{The Fermi surface of KCa$_2$Fe$_4$As$_4$O$_2$, each band is show separately.}
	\label{FS_bands}
\end{figure}

Also the Fermi surfaces at different values of electron (b-d) and hole (e-g) doping levels introduced by a rigid band shift are shown in Fig.~\ref{FS}.
The doping values correspond to Lifshitz transition or in other words appearance or disappearance of some Fermi surface sheets.
At electron doping level of 2.0 electrons per formula unit the O-2p Fermi surface sheet disappears, thus the Fermi surface of KCa$_2$Fe$_4$As$_4$O$_2$ system becomes similar to one of RbGd$_2$Fe$_4$As$_4$O$_2$~\cite{RbGd2Fe4As4O2_calc} or KCa$_2$Fe$_4$As$_4$F$_2$ system. For completeness, all 12 Fermi surface sheets are presented separately in Fig.~\ref{FS_bands} for stoichiometric  KCa$_2$Fe$_4$As$_4$O$_2$.

And finally, we established the magnetic ground-state by calculating the total energy of different magnetic phases (see Table~\ref{magnetic_phases}): paramagnetic phase (PM), ferromagnetic phase (FM), checkerboard antiferromagnetic phase (AFM1) (nearest-neighbor Fe spins antiparallel each other) and stripe antiferromagnetic phase (AFM2) (Fe spin moments parallel along one line of atoms and anti-parallel along neighbor line of atoms).
The magnetic ground state of KCa$_2$Fe$_4$As$_4$O$_2$ is found to be checkerboard antiferromagnetic (AFM1) one with $-41.1$~meV per Fe atom lower than the PM state.
\begin{table}
	\caption{The calculated total energy per Fe atom with respect to the paramagnetic phase (PM) and Fe magnetic moment of magnetic phases (AFM1 (Ch.B.) -- checkerboard antiferromagnetic order, AFM2 -- stripe antiferromagnetic order).}
\begin{tabular}{lll}
	\hline Order & Energy per Fe (meV) & Fe Magnetic Moment $\left(\mu_{B}\right)$ \\
	\hline PM & 0 & 0 \\
	AFM2 (stripe) & $-20.3$ & $1.04$ \\
	FM & $-27.6$ & $0.78$ \\
	AFM1 (Ch.B.) & $-41.1$ & $1.64$ \\
	\hline
\end{tabular}
\label{magnetic_phases}
\end{table}

\section{Conclusions}
In this paper we performed the DFT/GGA first-principle calculations of hole self-doped potential iron-based superconductor KCa$_2$Fe$_4$As$_4$O$_2$.
Most interesting result of our paper is presence at the Fermi level of anomalous metallic O-2p band which is formed by O-2p$_x$ and O-2p$_y$ orbitals.
The Fermi surface sheet corresponding of O-ps states has propeller shape and has its center at $\Gamma$-point.
The O-2p states are not filled completely. This tells us that the CaO layer gives some of the electrons to the FeAs layer.
The cylindrical Fe-3d Fermi surface sheets are modified due to hybridization with O-2p Fermi surface sheet.
Analysis of DOS obtained by the Wannier function projection provides self-doping value of Fe-As layer -- 0.75 hole/Fe-3d.
The similar analysis gives at O-2p states 5.5 electrons per O ion.
The magnetic ground state is the checkerboard antiferromagnetic state.
Also the comparative study of KCa$_2$Fe$_4$As$_4$O$_2$ with the same family systems KCa$_2$Fe$_4$As$_4$F$_2$ and RbGd$_2$Fe$_4$As$_4$O$_2$ was performed to prove that the KCa$_2$Fe$_4$As$_4$O$_2$ system is quite outstanding in this line.
We believe that KCa$_2$Fe$_4$As$_4$O$_2$ might be an potential superconductor with possible unusual properties because of anomalous metallic O-2p band.

\section*{Acknowledgements}
This work was supported in part by RSCF grant No. 21-12-00394.

\bibliography{./bib_file}

\begin{thebibliography}{10}
\expandafter\ifx\csname url\endcsname\relax
  \def\url#1{\texttt{#1}}\fi
\expandafter\ifx\csname urlprefix\endcsname\relax\def\urlprefix{URL }\fi
\expandafter\ifx\csname href\endcsname\relax
  \def\href#1#2{#2} \def\path#1{#1}\fi

\bibitem{KCaFe4As4_cryst_tr_Iyo2016}
A.~Iyo, K.~Kawashima, T.~Kinjo, T.~Nishio, S.~Ishida, H.~Fujihisa, Y.~Gotoh,
  K.~Kihou, H.~Eisaki, Y.~Yoshida,
  \href{https://pubs.acs.org/doi/10.1021/jacs.5b12571}{{New-Structure-Type
  Fe-Based Superconductors: CaAFe4As4 (A =K, Rb, Cs) and SrAFe4As4 (A = Rb,
  Cs)}}, Journal of the American Chemical Society 138~(10) (2016) 3410--3415.
\newblock \href {http://dx.doi.org/10.1021/jacs.5b12571}
  {\path{doi:10.1021/jacs.5b12571}}.
\newline\urlprefix\url{https://pubs.acs.org/doi/10.1021/jacs.5b12571}

\bibitem{KCaFe4As4_arpes_gap_Mou2016}
D.~Mou, T.~Kong, W.~R. Meier, F.~Lochner, L.-L. Wang, Q.~Lin, Y.~Wu, S.~L.
  Bud'ko, I.~Eremin, D.~D. Johnson, P.~C. Canfield, A.~Kaminski,
  \href{https://link.aps.org/doi/10.1103/PhysRevLett.117.277001}{{Enhancement
  of the Superconducting Gap by Nesting in CaKFe4As4: A New High Temperature
  Superconductor}}, Physical Review Letters 117~(27) (2016) 277001.
\newblock \href {http://dx.doi.org/10.1103/PhysRevLett.117.277001}
  {\path{doi:10.1103/PhysRevLett.117.277001}}.
\newline\urlprefix\url{https://link.aps.org/doi/10.1103/PhysRevLett.117.277001}

\bibitem{KCa2Fe4As4F2_lattice_Wang2016}
Z.~C. Wang, C.~Y. He, S.~Q. Wu, Z.~T. Tang, Y.~Liu, A.~Ablimit, C.~M. Feng,
  G.~H. Cao, {Superconductivity in KCa$_2$Fe$_4$As$_4$F$_2$ with separate
  double Fe$_2$As$_2$ layers}, Journal of the American Chemical Society
  138~(25) (2016) 7856--7859.
\newblock \href {http://dx.doi.org/10.1021/jacs.6b04538}
  {\path{doi:10.1021/jacs.6b04538}}.

\bibitem{RbGd2Fe4As4O2_exp}
Z.-C. Wang, C.-Y. He, S.-Q. Wu, Z.-T. Tang, Y.~Liu, A.~Ablimit, Q.~Tao, C.-M.
  Feng, Z.-A. Xu, G.-H. Cao,
  \href{https://iopscience.iop.org/article/10.1088/1361-648X/aa58d2}{{Superconductivity
  at 35 K by self doping in RbGd$_2$Fe$_4$As$_4$O$_2$}}, Journal of Physics:
  Condensed Matter 29~(11) (2017) 11LT01.
\newblock \href {http://dx.doi.org/10.1088/1361-648X/aa58d2}
  {\path{doi:10.1088/1361-648X/aa58d2}}.
\newline\urlprefix\url{https://iopscience.iop.org/article/10.1088/1361-648X/aa58d2}

\bibitem{BaTh2Fe4As4_NO2_exp}
Y.-t. Shao, Z.-c. Wang, B.-z. Li, S.-Q. Wu, J.-f. Wu, Z.~Ren, S.-w. Qiu,
  C.~Rao, C.~Wang, G.-H. Cao,
  \href{http://link.springer.com/10.1007/s40843-019-9438-7}{{BaTh2Fe4As4(N0.7O0.3)2:
  An iron-based superconductor stabilized by inter-block-layer charge
  transfer}}, Science China Materials 62~(9) (2019) 1357--1362.
\newblock \href {http://dx.doi.org/10.1007/s40843-019-9438-7}
  {\path{doi:10.1007/s40843-019-9438-7}}.
\newline\urlprefix\url{http://link.springer.com/10.1007/s40843-019-9438-7}

\bibitem{RbLn2Fe4As4O2_lat_tr_mag_Wang2017}
Z.~C. Wang, C.~Y. He, S.~Q. Wu, Z.~T. Tang, Y.~Liu, G.~H. Cao, {Synthesis,
  Crystal Structure and Superconductivity in Rb$Ln_2$Fe$_4$As$_4$O$_2$ ($Ln =
  $Sm, Tb, Dy and Ho)}, Chemistry of Materials 29~(4) (2017) 1805--1812.
\newblock \href {http://dx.doi.org/10.1021/acs.chemmater.6b05458}
  {\path{doi:10.1021/acs.chemmater.6b05458}}.

\bibitem{ALn2Fe4As4O2_exp}
S.-Q. Wu, Z.-C. Wang, C.-Y. He, Z.-T. Tang, Y.~Liu, G.-H. Cao,
  \href{https://link.aps.org/doi/10.1103/PhysRevMaterials.1.044804}{{Superconductivity
  at 33-37 K in $ALn_2$Fe$_4$As$_4$O$_2$ ($A = $K and Cs; $Ln =
  $lanthanides)}}, Physical Review Materials 1~(4) (2017) 044804.
\newblock \href {http://dx.doi.org/10.1103/PhysRevMaterials.1.044804}
  {\path{doi:10.1103/PhysRevMaterials.1.044804}}.
\newline\urlprefix\url{https://link.aps.org/doi/10.1103/PhysRevMaterials.1.044804}

\bibitem{RbGd2Fe4As4O2_calc}
Z.~Wang, G.~Wang, X.~Tian,
  \href{http://dx.doi.org/10.1016/j.jallcom.2017.03.017}{{Electronic structure
  and magnetism of RbGd2Fe4As4O2}}, Journal of Alloys and Compounds 708 (2017)
  392--396.
\newblock \href {http://dx.doi.org/10.1016/j.jallcom.2017.03.017}
  {\path{doi:10.1016/j.jallcom.2017.03.017}}.
\newline\urlprefix\url{http://dx.doi.org/10.1016/j.jallcom.2017.03.017}

\bibitem{KCa2Fe4As4F2_calc_Wang2016}
G.~Wang, Z.~Wang, X.~Shi,
  \href{http://link.springer.com/10.1007/s10948-019-05367-3
  https://iopscience.iop.org/article/10.1209/0295-5075/116/37003}{{Self-hole-doping–induced
  superconductivity in KCa$_2$Fe$_4$As$_4$F$_2$}}, EPL (Europhysics Letters)
  116~(3) (2016) 37003.
\newblock \href {http://dx.doi.org/10.1209/0295-5075/116/37003}
  {\path{doi:10.1209/0295-5075/116/37003}}.
\newline\urlprefix\url{http://link.springer.com/10.1007/s10948-019-05367-3
  https://iopscience.iop.org/article/10.1209/0295-5075/116/37003}

\bibitem{KCa2Fe4As4F2_calc_Singh2018}
\href{http://aip.scitation.org/doi/abs/10.1063/1.5052071}{{Density functional
  study of ACa$_2$Fe$_4$As$_4$F$_2$ (A = K, Rb): Electronic structure,
  unconventional superconductors}}, in: AIP Conference Proceedings, Vol. 2009,
  2018, p. 020002.
\newblock \href {http://dx.doi.org/10.1063/1.5052071}
  {\path{doi:10.1063/1.5052071}}.
\newline\urlprefix\url{http://aip.scitation.org/doi/abs/10.1063/1.5052071}

\bibitem{KCa2Fe4As4F2_calc_Li2020}
X.~Li, C.~Huang, Y.~Zhu, Y.~Zhang, {Magnetic Orders and Electronic Structures
  of Compressive- and Tensile-Strained KCa$_2$Fe$_4$As$_4$F$_2$ Films}, Journal
  of Superconductivity and Novel Magnetism 33~(5) (2020) 1377--1383.
\newblock \href {http://dx.doi.org/10.1007/s10948-019-05367-3}
  {\path{doi:10.1007/s10948-019-05367-3}}.

\bibitem{CsCa2Fe4As4F2_calc_Singh2018}
B.~Singh, P.~Kumar,
  \href{http://aip.scitation.org/doi/abs/10.1063/1.5033084}{{Unconventional
  iron-based superconductor CsCa$_2$Fe$_4$As$_4$F$_2$: A first-principle
  study}}, in: AIP Conference Proceedings, Vol. 1953, 2019, p. 120019.
\newblock \href {http://dx.doi.org/10.1063/1.5033084}
  {\path{doi:10.1063/1.5033084}}.
\newline\urlprefix\url{http://aip.scitation.org/doi/abs/10.1063/1.5033084}

\bibitem{KCa2Fe4As4F2_mag_Smidman2018}
M.~Smidman, F.~K.~K. Kirschner, D.~T. Adroja, A.~D. Hillier, F.~Lang, Z.~C.
  Wang, G.~H. Cao, S.~J. Blundell,
  \href{https://link.aps.org/doi/10.1103/PhysRevB.97.060509}{{Nodal multigap
  superconductivity in KCa2Fe4As4F2}}, Physical Review B 97~(6) (2018) 060509.
\newblock \href {http://dx.doi.org/10.1103/PhysRevB.97.060509}
  {\path{doi:10.1103/PhysRevB.97.060509}}.
\newline\urlprefix\url{https://link.aps.org/doi/10.1103/PhysRevB.97.060509}

\bibitem{KCa2Fe4As4F2_ARPES}
D.~Wu, W.~Hong, C.~Dong, X.~Wu, Q.~Sui, J.~Huang, Q.~Gao, C.~Li, C.~Song,
  H.~Luo, C.~Yin, Y.~Xu, X.~Luo, Y.~Cai, J.~Jia, Q.~Wang, Y.~Huang, G.~Liu,
  S.~Zhang, F.~Zhang, F.~Yang, Z.~Wang, Q.~Peng, Z.~Xu, X.~Qiu, S.~Li, H.~Luo,
  J.~Hu, L.~Zhao, X.~J. Zhou,
  \href{https://link.aps.org/doi/10.1103/PhysRevB.101.224508}{{Spectroscopic
  evidence of bilayer splitting and strong interlayer pairing in the
  superconductor KCa2Fe4As4F2 Dingsong}}, Physical Review B 101~(22) (2020)
  224508.
\newblock \href {http://dx.doi.org/10.1103/PhysRevB.101.224508}
  {\path{doi:10.1103/PhysRevB.101.224508}}.
\newline\urlprefix\url{https://link.aps.org/doi/10.1103/PhysRevB.101.224508}

\bibitem{KCa2Fe4As4F2_elast_resist_Terashima2020}
T.~Terashima, Y.~Matsushita, H.~Yamase, N.~Kikugawa, H.~Abe, M.~Imai, S.~Uji,
  S.~Ishida, H.~Eisaki, A.~Iyo, K.~Kihou, C.-H. Lee, T.~Wang, G.~Mu,
  \href{https://link.aps.org/doi/10.1103/PhysRevB.102.054511}{{Elastoresistance
  measurements on CaKFe4As4 and KCa2Fe4As4F2 with the Fe site ofC2v symmetry}},
  Physical Review B 102~(5) (2020) 054511.
\newblock \href {http://dx.doi.org/10.1103/PhysRevB.102.054511}
  {\path{doi:10.1103/PhysRevB.102.054511}}.
\newline\urlprefix\url{https://link.aps.org/doi/10.1103/PhysRevB.102.054511}

\bibitem{KCa2Fe4As4F2_wire}


\bibitem{KCa2Fe4As4F2_torque_Yu2019}
A.~B. Yu, T.~Wang, Y.~F. Wu, Z.~Huang, H.~Xiao, G.~Mu, T.~Hu,
  \href{https://link.aps.org/doi/10.1103/PhysRevB.100.144505}{{Probing
  superconducting anisotropy of single crystal KCa2Fe4As4F2 by magnetic torque
  measurements}}, Physical Review B 100~(14) (2019) 144505.
\newblock \href {http://dx.doi.org/10.1103/PhysRevB.100.144505}
  {\path{doi:10.1103/PhysRevB.100.144505}}.
\newline\urlprefix\url{https://link.aps.org/doi/10.1103/PhysRevB.100.144505}

\bibitem{KCa2Fe4As4F2_Hc2_Wang2019}
T.~Wang, J.~Chu, H.~Jin, J.~Feng, L.~Wang, Y.~Song, C.~Zhang, X.~Xu, W.~Li,
  Z.~Li, T.~Hu, D.~Jiang, W.~Peng, X.~Liu, G.~Mu, {Single-Crystal Growth and
  Extremely High Hc2 of 12442-Type Fe-Based Superconductor KCa2Fe4As4F2},
  Journal of Physical Chemistry C 123~(22) (2019) 13925--13929.
\newblock \href {http://dx.doi.org/10.1021/acs.jpcc.9b04624}
  {\path{doi:10.1021/acs.jpcc.9b04624}}.

\bibitem{KCa2Fe4As4F2_Hc2_Wang2020}
T.~Wang, C.~Zhang, L.~Xu, J.~Wang, S.~Jiang, Z.~Zhu, Z.~Wang, J.~Chu, J.~Feng,
  L.~Wang, W.~Li, T.~Hu, X.~Liu, G.~Mu,
  \href{http://link.springer.com/10.1007/s11433-019-1441-4}{{Strong Pauli
  paramagnetic effect in the upper critical field of KCa2Fe4As4F2}}, Science
  China Physics, Mechanics \& Astronomy 63~(2) (2020) 227412.
\newblock \href {http://dx.doi.org/10.1007/s11433-019-1441-4}
  {\path{doi:10.1007/s11433-019-1441-4}}.
\newline\urlprefix\url{http://link.springer.com/10.1007/s11433-019-1441-4}

\bibitem{KCa2Fe4As4F2_neutron_Hong2020}
W.~Hong, L.~Song, B.~Liu, Z.~Li, Z.~Zeng, Y.~Li, D.~Wu, Q.~Sui, T.~Xie,
  S.~Danilkin, H.~Ghosh, A.~Ghosh, J.~Hu, L.~Zhao, X.~Zhou, X.~Qiu, S.~Li,
  H.~Luo, \href{https://doi.org/10.1103/PhysRevLett.125.117002}{{Neutron Spin
  Resonance in a Quasi-Two-Dimensional Iron-Based Superconductor}}, Physical
  Review Letters 125~(11) (2020) 117002.
\newblock \href {http://dx.doi.org/10.1103/PHYSREVLETT.125.117002}
  {\path{doi:10.1103/PHYSREVLETT.125.117002}}.
\newline\urlprefix\url{https://doi.org/10.1103/PhysRevLett.125.117002}

\bibitem{KCa2Fe4As4F2_systematic_Pyon2020}
S.~Pyon, Y.~Kobayashi, A.~Takahashi, W.~Li, T.~Wang, G.~Mu, A.~Ichinose,
  T.~Kambara, A.~Yoshida, T.~Tamegai,
  \href{https://link.aps.org/doi/10.1103/PhysRevMaterials.4.104801}{{Anisotropic
  physical properties and large critical current density in KCa2Fe4As4 F2
  single crystal}}, Physical Review Materials 4~(10) (2020) 104801.
\newblock \href {http://dx.doi.org/10.1103/PhysRevMaterials.4.104801}
  {\path{doi:10.1103/PhysRevMaterials.4.104801}}.
\newline\urlprefix\url{https://link.aps.org/doi/10.1103/PhysRevMaterials.4.104801}

\bibitem{KCa2Fe4As4F2_press_Wang2019}
B.~Wang, Z.-C. Wang, K.~Ishigaki, K.~Matsubayashi, T.~Eto, J.~Sun, J.-G. Cheng,
  G.-H. Cao, Y.~Uwatoko,
  \href{https://link.aps.org/doi/10.1103/PhysRevB.99.014501}{{Pressure-induced
  enhancement of superconductivity and quantum criticality in the 12442-type
  hybrid-structure superconductor KCa2Fe4As4F2}}, Physical Review B 99~(1)
  (2019) 014501.
\newblock \href {http://dx.doi.org/10.1103/PhysRevB.99.014501}
  {\path{doi:10.1103/PhysRevB.99.014501}}.
\newline\urlprefix\url{https://link.aps.org/doi/10.1103/PhysRevB.99.014501}

\bibitem{KCa2Fe4As4F2_tr_Wang2020}
T.~Wang, J.~Chu, J.~Feng, L.~Wang, X.~Xu, W.~Li, H.~Wen, X.~Liu, G.~Mu,
  \href{https://link.springer.com/10.1007/s11433-020-1549-9}{{Low temperature
  specific heat of 12442-type KCa2Fe4As4F2 single crystals}}, Science China
  Physics, Mechanics \& Astronomy 63~(9) (2020) 297412.
\newblock \href {http://dx.doi.org/10.1007/s11433-020-1549-9}
  {\path{doi:10.1007/s11433-020-1549-9}}.
\newline\urlprefix\url{https://link.springer.com/10.1007/s11433-020-1549-9}

\bibitem{KCa2Fe4As4F2_NMR_Wang2017}
Z.~C. Wang, C.~Y. He, S.~Q. Wu, Z.~T. Tang, Y.~Liu, G.~H. Cao, {Synthesis,
  Crystal Structure and Superconductivity in RbLn2Fe4As4O2 (Ln = Sm, Tb, Dy,
  and Ho)}, Chemistry of Materials 29~(4) (2017) 1805--1812.
\newblock \href {http://dx.doi.org/10.1021/acs.chemmater.6b05458}
  {\path{doi:10.1021/acs.chemmater.6b05458}}.

\bibitem{KCa2Fe4As4F2_NMR_Luo2020}
J.~Luo, C.~Wang, Z.~Wang, Q.~Guo, J.~Yang, R.~Zhou, K.~Matano, T.~Oguchi,
  Z.~Ren, G.~Cao, G.-Q. Zheng,
  \href{https://iopscience.iop.org/article/10.1088/2053-1583/abe778
  https://iopscience.iop.org/article/10.1088/1674-1056/ab892d}{{NMR and NQR
  studies on transition-metal arsenide superconductors LaRu2As2, KCa2Fe4As4F2,
  and A2Cr3As3}}, Chinese Physics B 29~(6) (2020) 067402.
\newblock \href {http://dx.doi.org/10.1088/1674-1056/ab892d}
  {\path{doi:10.1088/1674-1056/ab892d}}.
\newline\urlprefix\url{https://iopscience.iop.org/article/10.1088/2053-1583/abe778
  https://iopscience.iop.org/article/10.1088/1674-1056/ab892d}

\bibitem{ACa2Fe4As4F2_lat_tr_mag_Wang2017}
Z.~C. Wang, C.~Y. He, S.~Q. Wu, Z.~T. Tang, Y.~Liu, G.~H. Cao, {Synthesis,
  Crystal Structure and Superconductivity in RbLn2Fe4As4O2 (Ln = Sm, Tb, Dy,
  and Ho)}, Chemistry of Materials 29~(4) (2017) 1805--1812.
\newblock \href {http://dx.doi.org/10.1021/acs.chemmater.6b05458}
  {\path{doi:10.1021/acs.chemmater.6b05458}}.

\bibitem{KCa2FeCo4As4F2_tr_mag_Ishida2017}
J.~Ishida, S.~Iimura, H.~Hosono,
  \href{https://link.aps.org/doi/10.1103/PhysRevB.96.174522}{{Effects of
  disorder on the intrinsically hole-doped iron-based superconductor
  KCa2Fe4As4F2 by cobalt substitution}}, Physical Review B 96~(17) (2017)
  174522.
\newblock \href {http://dx.doi.org/10.1103/PhysRevB.96.174522}
  {\path{doi:10.1103/PhysRevB.96.174522}}.
\newline\urlprefix\url{https://link.aps.org/doi/10.1103/PhysRevB.96.174522}

\bibitem{CsCa2Fe4As4F2_tr_mag_Wang2019}
Z.-C. Wang, Y.~Liu, S.-Q. Wu, Y.-T. Shao, Z.~Ren, G.-H. Cao,
  \href{https://link.aps.org/doi/10.1103/PhysRevB.99.144501}{{Giant anisotropy
  in superconducting single crystals of CsCa2Fe4As4F2}}, Physical Review B
  99~(14) (2019) 144501.
\newblock \href {http://dx.doi.org/10.1103/PhysRevB.99.144501}
  {\path{doi:10.1103/PhysRevB.99.144501}}.
\newline\urlprefix\url{https://link.aps.org/doi/10.1103/PhysRevB.99.144501}

\bibitem{CsCa2Fe4As4F2_tr_mag_Huang2019}
Y.~Y. Huang, Z.~C. Wang, Y.~J. Yu, J.~M. Ni, Q.~Li, E.~J. Cheng, G.~H. Cao,
  S.~Y. Li,
  \href{https://link.aps.org/doi/10.1103/PhysRevB.99.020502}{{Multigap nodeless
  superconductivity in CsCa2Fe4As4 F2 probed by heat transport}}, Physical
  Review B 99~(2) (2019) 020502.
\newblock \href {http://arxiv.org/abs/1811.06379} {\path{arXiv:1811.06379}},
  \href {http://dx.doi.org/10.1103/PhysRevB.99.020502}
  {\path{doi:10.1103/PhysRevB.99.020502}}.
\newline\urlprefix\url{https://link.aps.org/doi/10.1103/PhysRevB.99.020502}

\bibitem{RbCa2Fe4As4F2_Xing_2020}
X.~Xing, X.~Yi, M.~Li, Y.~Meng, G.~Mu, J.-Y. Ge, Z.~Shi,
  \href{https://doi.org/10.1088/1361-6668/abb35f}{Vortex phase diagram in
  12442-type {RbCa}2fe4as4f2 single crystal revealed by magneto-transport and
  magnetization measurements}, Superconductor Science and Technology 33~(11)
  (2020) 114005.
\newblock \href {http://dx.doi.org/10.1088/1361-6668/abb35f}
  {\path{doi:10.1088/1361-6668/abb35f}}.
\newline\urlprefix\url{https://doi.org/10.1088/1361-6668/abb35f}

\bibitem{WIEN2k2020}
P.~Blaha, K.~Schwarz, F.~Tran, R.~Laskowski, G.~K. Madsen, L.~D. Marks,
  \href{https://doi.org/10.1063/1.5143061}{{WIEN2k: An APW+lo program for
  calculating the properties of solids}}, The Journal of chemical physics
  152~(7) (2020) 074101.
\newblock \href {http://dx.doi.org/10.1063/1.5143061}
  {\path{doi:10.1063/1.5143061}}.
\newline\urlprefix\url{https://doi.org/10.1063/1.5143061}

\bibitem{PBE_1996}
J.~P. Perdew, K.~Burke, M.~Ernzerhof, {Generalized gradient approximation made
  simple}, Physical Review Letters 77~(18) (1996) 3865--3868.
\newblock \href {http://dx.doi.org/10.1103/PhysRevLett.77.3865}
  {\path{doi:10.1103/PhysRevLett.77.3865}}.

\bibitem{QE_Giannozzi2009}
\href{https://iopscience.iop.org/article/10.1088/0953-8984/21/39/395502}{{QUANTUM
  ESPRESSO: a modular and open-source software project for quantum simulations
  of materials}}, Journal of Physics: Condensed Matter 21~(39) (2009) 395502.
\newblock \href {http://dx.doi.org/10.1088/0953-8984/21/39/395502}
  {\path{doi:10.1088/0953-8984/21/39/395502}}.
\newline\urlprefix\url{https://iopscience.iop.org/article/10.1088/0953-8984/21/39/395502}

\bibitem{Wannier90_Pizzi2020}
G.~Pizzi, V.~Vitale, R.~Arita, S.~Bl{\"{u}}gel, F.~Freimuth, G.~G{\'{e}}ranton,
  M.~Gibertini, D.~Gresch, C.~Johnson, T.~Koretsune, J.~Iba{\~{n}}ez-Azpiroz,
  H.~Lee, J.-M. Lihm, D.~Marchand, A.~Marrazzo, Y.~Mokrousov, J.~I. Mustafa,
  Y.~Nohara, Y.~Nomura, L.~Paulatto, S.~Ponc{\'{e}}, T.~Ponweiser, J.~Qiao,
  F.~Th{\"{o}}le, S.~S. Tsirkin, M.~Wierzbowska, N.~Marzari, D.~Vanderbilt,
  I.~Souza, A.~A. Mostofi, J.~R. Yates, {Wannier90 as a community code: new
  features and applications}, Journal of Physics: Condensed Matter 32~(16)
  (2020) 165902.
\newblock \href {http://dx.doi.org/10.1088/1361-648x/ab51ff}
  {\path{doi:10.1088/1361-648x/ab51ff}}.

\bibitem{CaKFe4As4_Liu2020}
W.~Liu, L.~Cao, S.~Zhu, L.~Kong, G.~Wang, M.~Papaj, P.~Zhang, Y.-B. Liu,
  H.~Chen, G.~Li, F.~Yang, T.~Kondo, S.~Du, G.-H. Cao, S.~Shin, L.~Fu, Z.~Yin,
  H.-J. Gao, H.~Ding,
  \href{http://www.nature.com/articles/s41467-020-19487-1}{{A new Majorana
  platform in an Fe-As bilayer superconductor}}, Nature Communications 11~(1)
  (2020) 5688.
\newblock \href {http://dx.doi.org/10.1038/s41467-020-19487-1}
  {\path{doi:10.1038/s41467-020-19487-1}}.
\newline\urlprefix\url{http://www.nature.com/articles/s41467-020-19487-1}

\end{thebibliography}

\end{document}